\documentclass[a4paper,11pt]{article}
\pdfoutput=1 

\usepackage{jheppub} 

\usepackage[T1]{fontenc} 

\usepackage{cancel}
\usepackage{pifont}
\usepackage{amsthm}         
\usepackage{amsmath} 	   
\usepackage{amssymb}       
\usepackage{amsfonts}
\usepackage{graphicx} 	    
\usepackage{verbatim}
\usepackage{epsfig,bbm}
\usepackage{color}
\usepackage{colortbl}
\usepackage{float}
\usepackage{multirow}
\usepackage{booktabs}

\usepackage{subcaption}

\definecolor{babyblue}{rgb}{0.54, 0.81, 0.94}
\definecolor{corn}{rgb}{0.98, 0.93, 0.36}

\title{Dark Energy, Extra Dimensions, and the Swampland}

\author[a]{Gabriele Montefalcone,} 
\author[a,1]{Paul J. Steinhardt\note{Corresponding author.}}
\author[b]{and Daniel H. Wesley}

\affiliation[a]{Department of Physics, Princeton University, Princeton, NJ 08544, USA}
\affiliation[b]{PDT, 1745 Broadway, New York, NY 10019, USA}

\emailAdd{steinh@princeton.edu}

\abstract{Perhaps the greatest challenge for fundamental theories based on compactification from extra dimensions is accommodating a period of accelerated cosmological expansion.   Previous studies have identified constraints imposed by the existence of dark energy for two overlapping classes of compactified theories: (1)  those in which the higher dimensional picture satisfies certain metric properties selected to reproduce known low energy phenomenology; or (2)   those derived from string theory assuming they satisfy the Swampland conjectures. For either class, the analyses showed that dark energy is only possible if  it takes the form of quintessence. In this paper, we explore the consequences for theories that belong to both classes and show that the joint constraints are highly restrictive, leaving few options.}


\begin{document}
\maketitle
\flushbottom

\section{Introduction}

The notion that the fundamental laws of physics governing the observed universe  derive from the compactification of a general relativistic theory with extra dimensions has been a common element of many attempts at a unified theory, ranging from the seminal work of Kaluza \cite{Kaluza:1984ws} and Klein \cite{Klein:1926fj} to contemporary string theory.  The general concept has many aesthetically appealing aspects, and one might imagine that there are far too many degrees of freedom and parameters to rule out the entire spectrum of possibilities empirically.  

But maybe not. Over the last two decades, it has become increasingly clear that  a single observation -- the discovery of dark energy and cosmic acceleration -- poses a serious challenge.  Now large classes of extra-dimensional theories can be ruled out independent of the compactification scale.   

One line of argument \cite{Oxidised,Steinhardt:2008nk} assumes the metric of the extra-dimensional theory possesses properties (such as Ricci flatness) commonly selected to reproduce low energy particle and cosmology phenomenology. 
 We will refer to the conditions obtained from this line of argument as the {\it metric-based constraints.}  The key observation is that the accelerated expansion of the large three spatial dimensions causes the compactified extra-dimensions to vary with time, imposing constraints on   the equation
of state, $w_{DE}(z)$, where $z$ is the redshift and $w_{DE}$ is the ratio of the dark energy pressure to energy density \cite{Steinhardt:2008nk}.  
 
 The other line of argument applies specifically to string theory and results in the  {\it Swampland-based constraints} \cite{Vafa:2005ui,Ooguri:2006in,Obied:2018sgi}; see~\cite{Brennan:2017rbf,Palti:2019pca} for recent reviews.  Assuming that dark energy is due to a scalar field $\phi$ with positive potential energy density $V(\phi)>0$, the Swampland constraints impose conditions on the field range ($\Delta \phi$), the  potential slope ($\nabla_{\phi} V$) and curvature ($\nabla^{2}_{\phi} V$).
 
The two sets of constraints are  closely related in that $\nabla_{\phi} V/V$ and $\nabla^2_{\phi}  V/V$ are  directly  related to the equation of state $w_{DE}(z)$ and its time derivative in the limit that dark energy is due to a slowly-rolling scalar field.  (Here and throughout the paper,  reduced Planck units with $8 \pi G =1$ are assumed.)   
But the two types of constraints are also different in several important respects. 

The metric-based ones are mathematically rigorous  for any extra-dimensional theory satisfying the specific assumptions about the metric described in Sec.~\ref{AMetric} below. They are more general in that they do not assume the dark energy is due to a scalar field; rather, the constraints apply directly to $w_{DE}(z)$ independent of the microscopic source of dark energy.  The constraints are also quantitatively precise, providing specific numerical bounds on $w_{DE}(z)$, as illustrated in the figures of Ref.~\cite{Steinhardt:2008nk}.  

By contrast, the Swampland-based constraints assume string theory and  scalar field dark energy specifically.  The constraints are generally not as quantitatively precise as the metric constraints; for example, they are often expressed as the condition that certain quantities are ${\cal O}(1)$, where some discussions allow significant leeway in what this means numerically (though recently proposed constraints based on generalizations of the Transplanckian Censorship Conjecture are more stringent~\cite{Bedroya:2019snp,Andriot:2020lea}).   
At the same time, string theory may include self-consistent examples that do not satisfy the metric assumptions and so are not subject to the metric-based constraints.  In other words, the two classes of constraints apply to two  sets of extra-dimensional theories that are not identical but have enormous overlap, where that overlap includes commonly used examples in the literature. 

 In this paper, we confine ourselves to theories in the overlap and explore the consequences when both metric-based and Swampland-based constraints are imposed.  We show that the combination of constraints cannot be satisfied while remaining with current cosmological observations.  The remaining options for incorporating dark energy in compactified theories require evading at least some of the constraints.  These options are less explored and appear to be more complicated, as we discuss in the concluding section. 

\section{Assumptions underlying metric-based constraints}
\label{AMetric}
The metric-based constraints assume: (1)  the higher dimensional theory and the compactified theory, are described by  a		$(D+1)$-dimensional and $(3+1)$-dimensional Einstein-Hilbert action, respectively; (2)  the $(3+1)$-dimensional metric is spatially flat, in accordance with observations \cite{Aghanim:2018eyx}; and, (3) the extra $k=D-3$ dimensions are bounded and either Ricci-flat (RF) or Conformally Ricci-flat (CRF).  That is, the metric can be expressed as:
    \begin{equation}
        ds^2= e^{2\Omega(t,y)}g_{ \mu \nu}^{\text{FRW}}(t,x)dx^{\mu}dx^{\nu}+e^{-2\Omega(t,y)}\bar{h}_{ \alpha
        \beta}^{RF}(t,y)dy^{\alpha}dy^{\beta}
    \end{equation}
    where  $g_{\rm \mu \nu}^{\text{FRW}} $ is the flat Friedmann-Robertson Walker metric with scale factor $\bar{a}(t)$;
 $\mu$, $\nu$ are the indices along the 4 large dimensions with coordinates $x^{\mu}$; and $\alpha$,$\beta$ are the indices along the $k$ compact extra dimensions with coordinates $y^{\alpha}$. Finally the extra dimensional metric 
 \begin{equation}   
    h_{ \alpha \beta}\equiv e^{-2\Omega(t,y)}\bar{h}_{ \alpha \beta}^{RF}(t,y) 
 \end{equation}       
        is chosen such that $\bar{h}_{ \alpha
        \beta}^{RF}(t,y)$  has vanishing Ricci scalar curvature with warp factor $\Omega$ either constant in the RF case or temporally and spatially dependent in the CRF case.   
    
 The metric conditions 
correspond to common constructions of extra-dimensional theories in the literature.  Theories satisfying the RF condition
include the original Kaluza-Klein model, the Randall-Sundrum models, 
all one-dimensional
manifolds, $S^1/Z_2$  orbifolds as in braneworld models \cite{Randall:1999vf,Randall:1999ee,Horava:1995qa}, 
flat tori,
tori with nonnegative Ricci scalar, and
manifolds of exactly $\text{SU}(n)$, $\text{Sp}(n)$, $G2$ and $\text{Spin}(7)$ holonomy.
 Note that this class of metrics includes the Calabi-Yau and $G2$ holonomy manifolds commonly studied in string compactifications.
The CRF metric appears in warped Calabi-Yau \cite{Giddings:2001yu}  and warped conifold \cite{Klebanov:2000hb} constructions
( sometimes referred to as conformally Calabi-Yau metrics).

To describe a spatially-flat FRW spacetime after dimensional 
reduction, the metric 
$h_{\alpha \beta}(t,y)$ and warp function $\Omega(t,y)$ must be functions of 
time 
$t$ and the 
extra-dimensional coordinates $y^m$ only.   Following the convention 
in 
Ref.~\cite{Oxidised}, we parameterize the rate 
of change of 
$h_{\alpha \beta}$  using quantities $\xi$ and $\sigma_{mn}$ defined by
\begin{equation}
\label{xivar}
\frac{1}{2} \frac{d \, h_{\alpha \beta}}{d \, t} = \frac{1}{k} \xi h_{\alpha \beta} + 
\sigma_{\alpha \beta}
\end{equation}
where $h_{\alpha \beta} \sigma_{\alpha \beta} =0$ and where $\xi$ and  $\sigma$ are 
functions of time 
and the extra dimensions; this relation assumes the gauge
choice discussed in Ref.~\cite{Oxidised}.

Without loss of generality, we can take  
the space-space 
components of the Einstein-frame energy-momentum tensor $T_{MN}$ (where ${M,N}$ span all $k+4$ dimensions) 
to be  block 
diagonal with a 
$3 \times 3$ block describing the energy-momentum in the 
three non-compact 
dimensions and a $k \times k$ block for the $k$ compact directions;  
the 0-0 component 
is the higher dimensional energy density $\rho$.   

Associated with the two blocks of space-space components of $T_{MN}$ 
are two trace 
averages:
\begin{equation}
p_3 \equiv \frac{1}{3} \gamma_3^{\alpha \beta} T_{\alpha \beta}
 \; \; {\rm and} \; \;
p_k \equiv \frac{1}{k} \gamma_k^{\alpha \beta} T_{\alpha \beta},
\end{equation}
where $\gamma_{3,k}$ are respectively
the $3\times 3$ and $k \times k$ blocks of the higher
dimensional space-time metric.

The metric-based constraints are expressed in terms of 
$A$-averaged quantities \cite{Oxidised,Steinhardt:2008nk}:
\begin{equation}
\langle Q \rangle_A \; = 
\left( \int Q \, e^{A\Omega} \sqrt{g} \; \text{d}^k y \right) {\Large 
/}
\left( \int e^{A\Omega} \sqrt{g} \; \text{d}^k y \right);
\end{equation}
that is, quantities averaged over the extra dimensions with weight 
factor $e^{A \Omega}$ where $A$ is a constant.  

For  simplicity, we will restrict ourself henceforth to the case of the CRF metric; the RF case is similar.  Terms in the Einstein equations  dependent on $\bar{a}$ can be expressed
in terms 
of the 
 4d effective scale factor using the relation \cite{Oxidised,Steinhardt:2008nk}: 
$a(t) \equiv 
e^{\phi/2}\bar{a}(t)$, where 
\begin{equation}\label{phieq}
e^{\phi} \equiv
 \int e^{2\Omega} \sqrt{g} \; \text{d}^k y.
\end{equation}
Then the 4d effective scale factor $a(t)$ obeys 
 the usual 4d Friedmann 
equations:
\begin{align}
\left( \frac{\dot{a}}{a} \right)^2 & = \frac{1}{3} \rho_{4d} \\
\left(\frac{\dot{a}}{a}\right)^2 + 2 \frac{\ddot{a}}{a}&=- p_{4d}.
\end{align}
Note that the 4d effective  energy density  $\rho_{4d}$ and pressure 
$p_{4d}$ are 
generally different from $\rho$ and $p_3$ in the higher dimensional 
theory if the warp factor $\Omega$ is non-trivial.  
Then, after substituting the CRF metric in  the Einstein equations, we obtain two conditions:
\begin{align}
\label{eq:QuinRhoPlusP3}
e^{-\phi}
\langle  e^{2\Omega} (\rho + p_3) \rangle_A   &=  (\rho_{4d} +p_{4d})
- \frac{k+2}{2k} \langle \xi \rangle_A^2   \notag \\  &- \frac{k+2}{2k} \langle (\xi 
- \langle 
\xi\rangle_A)^2\rangle_A \;  - \langle \sigma^2 \rangle_A \; \\
\label{Astar}
e^{-\phi}
\langle e^{2\Omega} (\rho + p_k) \rangle_A \; &= 
\frac{1}{2}\left( \rho_{4d} +3p_{4d} \right) 
+2(\frac{A}{4}-1)\frac{k+2}{2k} \langle (\xi - \langle \xi\rangle_A)^2\rangle_A
\; \notag \\
&  - \frac{k+2}{2k} \langle \xi \rangle_A^2 
  - \langle \sigma^2 \rangle_A\; \notag \\
&  + \left[ -5 + \frac{10}{k} +k+A(-3+\frac{6}{k}) \right]
\langle e^{2\Omega} (\partial\Omega)^2 \rangle_A\; \notag \\
&  + \frac{k+2}{2k} \frac{1}{a^3} \,\frac{d}{d t} \left(a^3 \langle
\xi \rangle_A  \right)
\end{align}
which  can be rewritten in a form that will be most convenient for analysis:
\begin{align}
\label{essentials1}
e^{-\phi}
\langle  e^{2\Omega} (\rho + p_3) \rangle_A \;  &   =  
\rho_{4d}(1  +w)
- \frac{k+2}{2k} \langle \xi \rangle_A \;^2. \notag \\ &  +               
{\rm non-positive \, terms \, for \, {\it all} \, A }
\\  \label{essentials2}
e^{-\phi}
\langle e^{2\Omega} (\rho + p_k) \rangle_A\; & =  
\frac{1}{2}\rho_{4d}(1+3 w) + \frac{k+2}{2k} \frac{1}{a^3} 
\,\frac{d}{d t} 
\left(a^3 \langle \xi \rangle_A  \right) \notag \\
&  +{\rm non-positive \, terms \, for \, 4>A>A*},
\end{align}
where $A*=4/3$ for $k=6$ (the case relevant to string theory) and the last term is precisely zero for $A=A*$.
In these expressions,
$w$ represents the ratio of the 
total 4d effective 
pressure $p_{4d}$ to the total 4d effective energy density 
$\rho_{4d}$.  

\section{Assumptions underlying the Swampland-based constraints}
\label{Swamp}

The Swampland of string theory is comprised of the subset of `consistent looking' $(3+1)$-dimensional effective quantum field theories coupled to gravity that are also consistent with  string theory~\cite{Vafa:2005ui,Ooguri:2006in,Obied:2018sgi}; for recent
reviews see~\cite{Brennan:2017rbf,Palti:2019pca}.  It has been conjectured that members of this subset satisfy the following conditions:

\noindent
{\sc Range condition}:  {\it  The range traversed by scalar fields in field space is
  bounded by $\Delta \sim {\cal{O}}(1)$}
\cite{Ooguri:2006in}.  More precisely, consider a theory of quantum
gravity coupled to a number of scalars $\phi^i$ in which
the effective Lagrangian, valid up to a cutoff scale $\Lambda$, takes the form
\begin{align}
  \label{lagrange}
  \mathcal{L}
  &=
  \sqrt{|g|} 
  \left[\frac{1}{2} R
    -\frac{1}{2}
    g^{\mu\nu}\partial_{\mu} \phi^i \partial_{\nu} \phi^j
    G_{ij}(\phi)
    -V(\phi)
    +\ldots
  \right],
\end{align}
  Note that, by expressing the theory in Einstein frame as above, 
 $G_{ij}(\phi)$ 
defines a field metric which we use to measure distances in the field space
$\phi^i$.
Then, it is conjectured,  there
is a finite radius $\cal{O}$(1) in field space where the effective Lagrangian above
is valid. 
See in
particular~\cite{Grimm:2018ohb,Heidenreich:2018kpg, Blumenhagen:2018hsh}
for a recent discussion and extensions,
of this conjecture.  Without loss of generality we will only consider  for the remainder of this paper  cases of dark energy due to a single scalar field $\phi$ and potential $V(\phi)$.  

The second Swampland criterion, which was
first conjectured in \cite{Obied:2018sgi}, is motivated by failed attempts
 to construct   dS or nearly-dS vacua in string theoretic models in a controlled approximation: \\
{\sc Slope condition}:  {\it For all $\phi$ for which $V(\phi)>0$,   $|\nabla_{\phi} V|/V
>c $ where $c>0$ is ${\cal O}$(1).}

Later a ``refined'' conjecture was introduced that allowed a second option \cite{Garg:2018reu,Ooguri:2018wrx}: the slope could be zero or much less than unity provided the curvature satisfies  $min(\nabla^2_{\phi} V)< -c' V$ where  $c'>0$ is  ${\cal O}$(1).   For the purposes of this paper, though, we can ignore this or any other Swampland conjectures that enable near-zero slope because we are  considering the specific application of scalar fields as models for dark energy.    A dark energy phase in which the scalar field rests at  a local or global maximum with zero slope and negative curvature has $w_{DE} \approx -1$, which is ruled out under the metric assumptions described above.\cite{Oxidised,Steinhardt:2008nk}   Complex scenarios in which the field is slowly rolling near a ``hilltop'' or experiences a turning point are ruled out by a more detailed dynamical analysis \cite{Obied:2018sgi}.  Roughly speaking, in these models the kinetic energy of the field scales so rapidly with redshift (that is, going back in time) that it overtakes the matter and radiation density, leading to unacceptable deviations from standard big bang expansion in the past, including violations of large scale structure and nucleosynthesis constraints. 
Since we are explicitly considering in this paper models that satisfy {\it both} the metric- and Swampland-based constraints as well as observational constraints, only the Swampland condition $|\nabla_{\phi} V|/V> {\cal O}(1) $ is relevant.

\section{Dark energy and extra dimensions}

For theories that satisfy both the metric-based {\it and} Swampland-based assumptions, it is possible to make some surprisingly strong statements.  For example: cosmic acceleration is impossible for any  compactified theory in which: \begin{itemize}
\item[$i$] the size of the extra dimensions is fixed ($\xi=0$); 
\item[$ii$] the NEC is satisfied; and, 
\item[$iii$] the RF or CRF metric assumptions described in Sec.~\ref{AMetric} (and commonly assumed in many phenomenological applications) are satisfied.   
\end{itemize}

The proof is simple: For any degree of cosmic acceleration, it is impossible to satisfy these three conditions and also Eq.~(\ref{essentials2}).  To satisfy the NEC, the value of $\rho+p_k$ must be non-negative at all space-time points, so its average value over the internal dimensions averaged over any positive definite measure ({\it e.g.}, the left hand side of Eq.~(\ref{essentials2})), must be positive.  But cosmic acceleration corresponds to $w<-1/3$, in which case the right hand side of Eq.~(\ref{essentials2}) is negative if the three conditions above are satisfied.  

{\it Hence, at least one of the three conditions $i-iii$ has to give.  This is the essence of the metric-based constraints.}

\begin{figure}
    \centering
    \includegraphics[width=.90\linewidth]{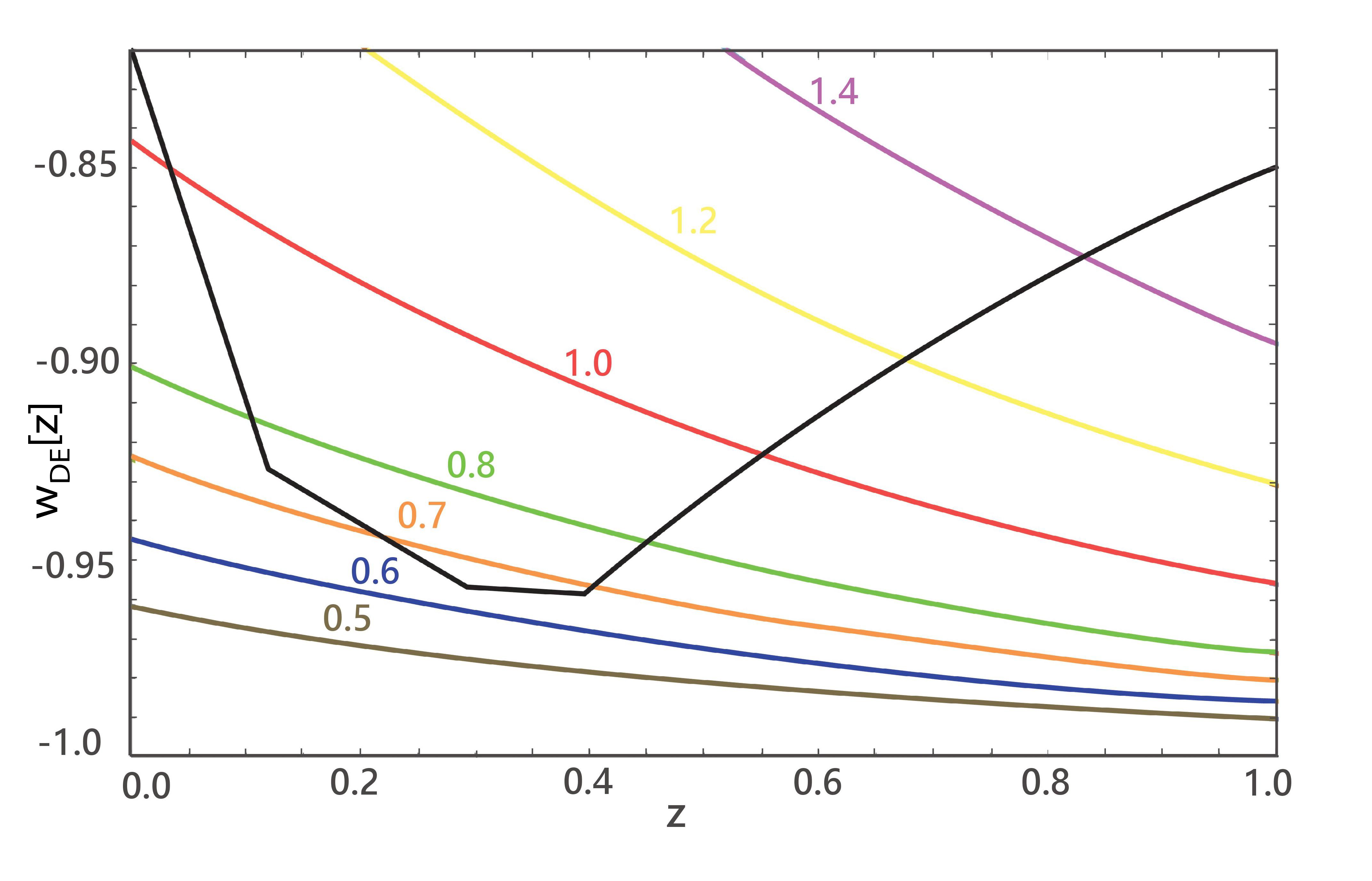}
    \caption[]{The black curve shows the current observational $2\sigma$ bound on $w_DE(z)$ for redshifts $0 < z < 1$ based on SNeIa, CMB and BAO
data \cite{Scolnic:2017caz}. This is compared with the predicted $w_{DE}(z)$ for exponential quintessence potentials with different values of constant
$\lambda$ under the constraint that $\Omega_{DE}(z = 0) = 0.7$ and assuming  $\Omega_{DE}(z)$ becomes negligible at $z > 1$. From the plot, it is clear that the upper bound is $\lambda \sim 0.6$ (blue curve).}
    \label{fig:1}
\end{figure}

This conclusion applies to string theories satisfying the Swampland-based constraints that also satisfy the three conditions above.  In Ref.~ \cite{Agrawal:2018own},  it was shown that the Swampland-based constraints rule out a cosmological constant or time-independent dark energy ($w_{DE}=-1$) and only permit quintessence dark energy \cite{Caldwell:1997ii} described by a scalar field $\phi$ rolling down a potential $V(\phi)$  with $V'/V = {\cal O}(1)$ or $w_{DE} \approx (1/3)(V'/V)^2 -1 = {\cal O}(-2/3)$.  As the authors showed, some leeway in what ${\cal O}(1)$ means precisely in the Swampland conjectures is required to satisfy current observational constraints.  In fact, models with the largest possible $V'/V$ that satisfy current observational constraint on dark energy were shown to be of the form:
\begin{equation}
V(\phi)=V_{\rm 0}e^{\lambda \phi}
    \label{eq:ST:3}
\end{equation}
for $\lambda \approx 0.6$ and with initial conditions given by the current constraints on $\Omega_{DE}(z)$ and $w_{ DE}(z)$ from supernovae (SNeIa), cosmic microwave background
(CMB) and baryon acoustic oscillation (BAO) measurements \cite{Scolnic:2017caz}.  Fig.\ref{fig:1} is a variant of Fig.~1 from Ref.~\cite{Agrawal:2018own} showing the constraints on the past evolution of $w_{DE}(z)$.

Fig.~2  shows the total cosmic equation of state ({\it i.e.,} the weighted sum over matter and dark energy) $w(a)\equiv \Omega_{DE}(a) w_{DE}(a) $ as a function of the scale factor $a= 1/(1+z)$, where $a=1$ today.  This figure shows both the past ($a<1$) and future ($a>1$) evolution of $w(a)$.

\begin{figure}
    \centering
    \includegraphics[width=.85\linewidth]{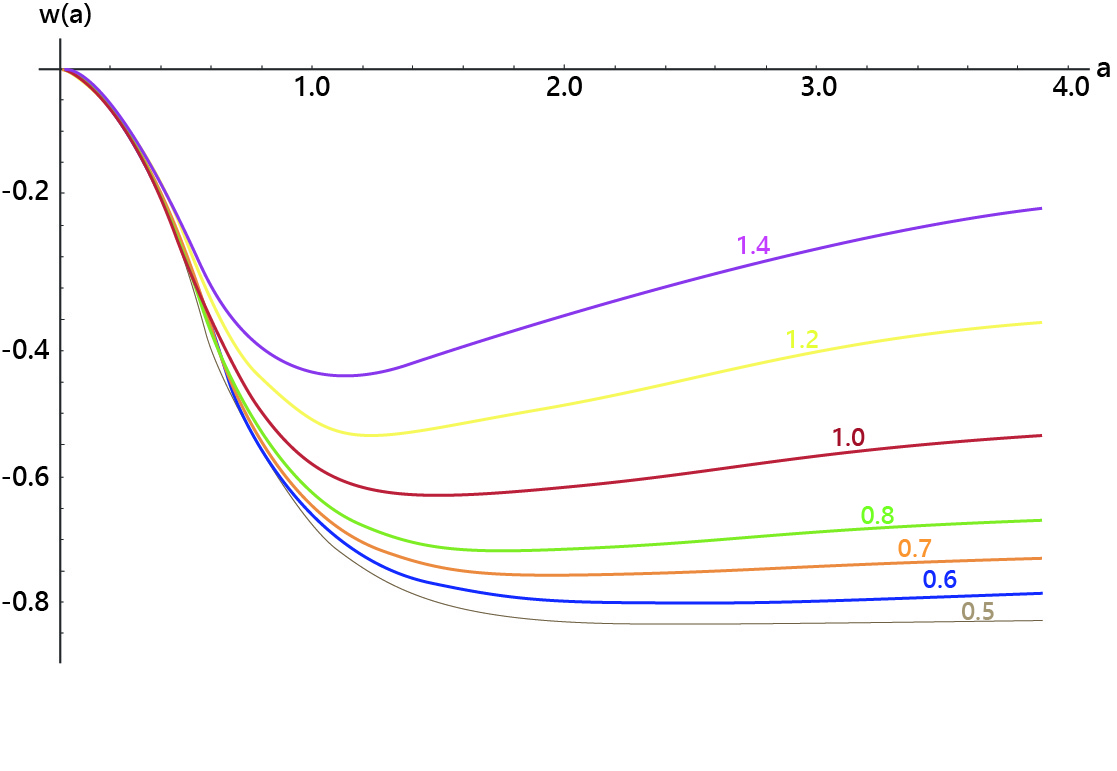}
    \caption[]{This plot shows a comparison the total cosmic equation of state $w(a)=w_{\rm DE}(a) \times \Omega_{\rm DE}(a)$ for $0.5 \leq \lambda \leq 1.4$ for scalar field dark energy models with exponential potentials, as motivated by considerations of Swampland-based constraints \cite{Agrawal:2018own}. }
    \label{fig:2}
\end{figure}

These stringy models nominally satisfy the Swampland-based constraints (given some leeway in the precise meaning of ${\cal O}(1)$) and current observational constraints on cosmic acceleration.   Of course, they are also derived by compactification from extra-dimensions, and so it is reasonable to consider the subset that also  satisfy the metric-based constraints (condition $iii$ above). 
 
An immediate conclusion is that models that satisfy both the Swampland-based and metric-based constraints  must 
violate either condition $i$ (fixed extra dimensions) and/or condition $ii$ (NEC).  In the following subsections, we consider the consequences of each of the two options.

\subsection{Option 1: Varying extra dimensions}

The time variation of the variable $\xi$ in the expression for $d h_{\alpha \beta}/dt$ in Eq.~\ref{xivar} determines the local expansion of the extra-dimensions. As shown in Ref.~\cite{Steinhardt:2008nk}, it is possible to have cosmic acceleration for a finite period and satisfy conditions $ii$ and $iii$  if $\xi$ is near zero and $d \xi/dt$ is large and positive.

For a  compactified 4d universe with Hubble parameter $H$, the variable  
\begin{equation}
    \zeta=\frac{1}{H}\int e^{2 \Omega} \xi \sqrt{h}d^ky,
    \label{eq:2.1}
\end{equation}
 is the extra-dimensional volume, where the weight factor (corresponding to $A=2$)  has been properly chosen such that $\zeta$ determines the four-dimensional Planck mass in warped compactifications.  Consequently, 
 $\zeta$ determines the variation of Newton's constant $G$:
\begin{equation}
    \frac{\Dot{G}}{G}= -H\zeta
    \label{eq:2.2}
\end{equation}

Then condition $ii$ (satisfying the NEC) corresponds to requiring the left hand sides of Eqs.~\ref{essentials1} and~\ref{essentials2} to be non-negative, or, equivalently, 
\begin{equation}
    \zeta^2\leq \frac{9(1+w)}{2}
    \label{eq:13}
   \end{equation}
\begin{equation}
  \frac{d\zeta}{dN}\geq \zeta^2+\frac{3(w-1)}{2}\zeta-\frac{9(1+3w)}{4}
   \label{eq:12}
\end{equation}
where 
\begin{equation} w(z)=p_{4d}/\rho_{4d} = \Omega_{DE}(z) w_{DE}(z)\end{equation} 
is the total cosmic equation of state, $N=\ln(a)$ and $a$ is the Einstein frame scale factor and $k=6$ was chosen because the intended application is to string theory.  (Note that possible contributions from $\langle \sigma^2 \rangle_{\rm A}$ and $\langle e^{2\Omega}(\partial\Omega)^2\rangle_{\rm A}$ are  take into account by the inequalities since they only make non-positive contributions to the right hand side of Eqs.~\ref{essentials1} and~\ref{essentials2} that make the NEC more difficult to satisfy.) 

\begin{figure}
    \centering
    \includegraphics[width=.95\linewidth]{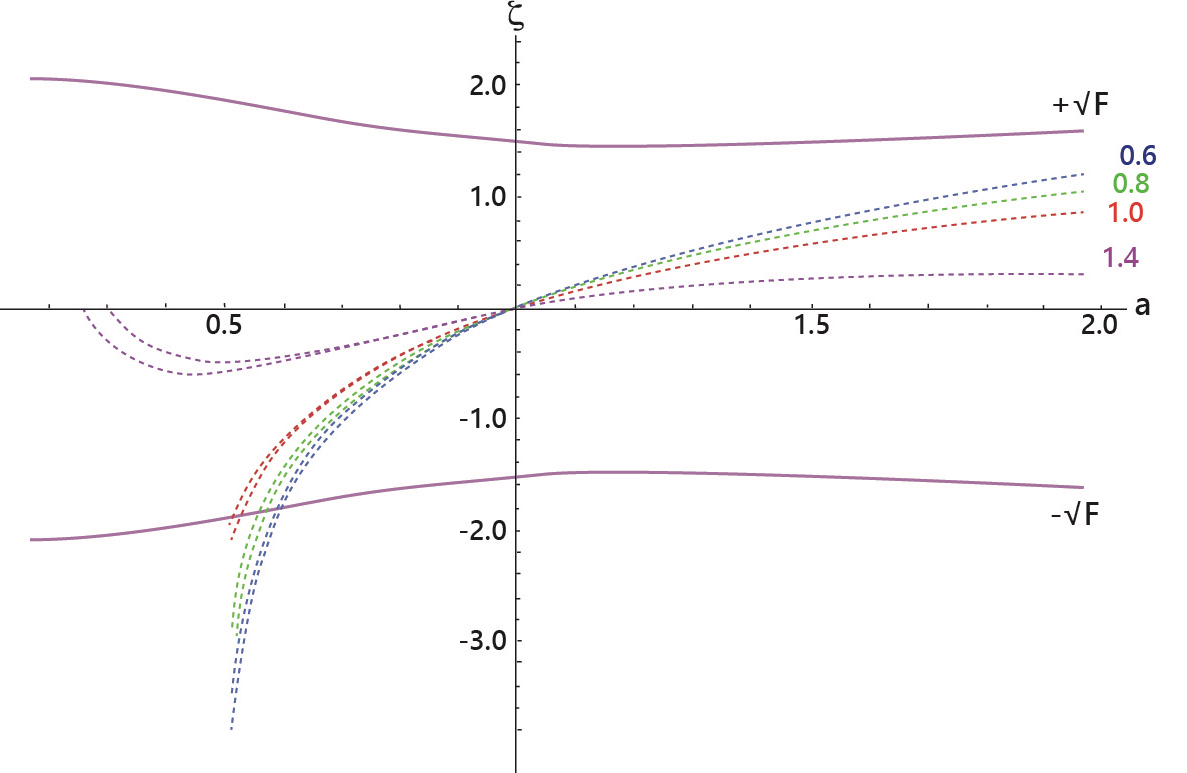}
    \caption[]{The dotted curves represent the trajectories of  $\zeta(a) \equiv \dot{G}/GH$  for scalar field dark energy with potential $V(\phi) = V_0 e^{\lambda \phi}$, where the value of $\zeta(a=1)$ is fixed so that $\dot{G}/GH$ today is consistent within $3 \sigma$ of the current limits on the instantaneous variation of $G$; see Eq.~\ref{lim}.  The curves labeled $\pm \sqrt{F}$  correspond to the NEC violating boundaries; that is, $\zeta$-trajectories that cross these curves must violate the null energy conditions in addition to having time-varying $G$ and time-varying extra-dimensions.    Trajectories are shown for $\lambda=0.6;0.8;1.0;1.4$.   These all correspond to models that satisfy the Swampland-based constraints; but only the case with $\lambda \le 0.6$ also satisfies the observational constraints on $w_{DE}(z)$ shown in Fig.~\ref{fig:1}.}
    \label{fig:4}
\end{figure}

Eq.~\ref{eq:13} is the condition that $\zeta(a)$  must obey in order for the NEC (condition $ii$ above) to be satisfied, as we are requiring in this subsection.  The condition is equivalent to demanding that $\zeta(z)$ remain  between the two outer curves in Fig.~\ref{fig:4} equal to $\pm \sqrt{F}$ where $F= \frac{9(1+w)}{2}$.

We can compute the  trajectories which satisfy NEC with minimal Newton constant variation by saturating the inequality in 
eq.~\ref{eq:12}. More precisely, the instantaneous variation in $G$ today is given by
\begin{equation}
\frac{\Dot{G}}{G}\Bigr|_{\substack{\text{today}}}=-H_{\rm 0}  \, \zeta(a=1)  
\end{equation}
and the secular variation in $G$ since big bang nucleosynthesis is given by
\begin{equation}
 \frac{G_{\rm BBN}}{G}=\text{exp} \; \Big[ \int^1_{a_{\rm BBN}} \frac{1}{a} \zeta(a) da \Big]
\label{eq:CCC}
\end{equation}
(Note that, in discussing $\Dot{G}/G$, we switch to the convention of observers who report constraints after first transforming to the Jordan frame.)

In Ref.~ \cite{Exploring-ExtraD}, it appeared that there exists a finite set of trajectories $\zeta(a)$ that could satisfy both the instantaneous and secular observational constraints on the variation of $G$ while remaining between the two outer curves forever to the past and substantially to the future (and thereby enabling the NEC to be satisfied).

Since then, the observational limits on secular variation have improved to
\begin{equation}
\frac{G_{\rm BBN}}{G} = 0.98 \pm 0.06
\end{equation}
at the 95\% confidence level~\cite{Alvey:2019ctk}, and the observational limits on instantaneous variation of $G$ today have been reduced by roughly two orders of magnitude due to improvements in the ephemeris of Mars together with improved data and modeling of the effects of the asteroid belt \cite{will2018theory}:
\begin{align}
    \frac{\Dot{G}}{G}\Bigr|_{\substack{\text{today}}} &= (0.00014 \pm 0.002) H_{\rm 0}  \\& \simeq (0.143 \pm 2.04)\times 10^{-13} h \, \text{yr}^{-1}
    \label{C:GdotNEW}.
\end{align}
where $H_0=100 h$ and $h \approx 0.7$ according to current measurements \cite{Aghanim:2018eyx}. 

Now we combine these metric-based constraints on dark energy with the Swampland-based constraints by computing the $\zeta(a)$ trajectories for quintessence scalar field dark energy with potential $V(\phi) = V_0 e^{\lambda \phi}$ where $\lambda = {\cal O}(1)$ according to the Swampland slope constraint.   For each $\lambda$, we constrain the instantaneous value of $\dot{G}/G$ to lie within the current $3 \sigma$ limits, or, equivalently, 
\begin{equation}
\label{lim}
\zeta(a=1) =0.00014 \pm 0.006
\end{equation}

Fig.~\ref{fig:4} shows the results: Except for $\lambda=1.4$, which is grossly inconsistent with observational limits on $w_{DE}(z)$, as shown in Fig.~\ref{fig:1}, all the curves that satisfy the Swampland-based based constraints have $\zeta$-trajectories that dive down to large negative values when extrapolated back in time.  That means these models violate  the $\frac{G_{\rm BBN}}{G}$ constraint and are unable to avoid entering the NEC-violating region.  This includes the case of $\lambda \le 0.6$.  

(For $\lambda=1.4$, it is possible to have $\zeta=0$ between $a=0$ and $a\approx 0.3$, and then have $\zeta$ deviate away from zero in such a way that the secular and instantaneous constraints on $G$ variation can be satisfied.  However, these models cannot produce the observed acceleration or the observed dark energy equation of state)

Hence, we conclude that all models that satisfy both the metric-based and Swampland based constraints and also satisfy the NEC necessarily violate current observational constraints on the dark energy equation of state and the variation of $G$. That is, violating only condition $i$ above is not sufficient for obtaining an acceptable model.

\subsection{Option 2: Violating the NEC}

Violating the NEC makes it possible to satisfy both the higher-dimensional and compactified Einstein equations (Eqs.~\ref{essentials1} and~\ref{essentials2}) without varying the size of the extra dimensions ($\xi=0$); that is, with NEC violation, it is possible to add positive contributions to the right hand sides of either or both equations so that they can be satisfied for substantial periods of cosmic acceleration despite keeping $G$ fixed.

In Ref.~\cite{Oxidised, Steinhardt:2008nk}, it was shown that compactified models satisfying the metric-based constraints and fixed extra dimensions (condition $i$) can  sustain periods of cosmic acceleration ({\it e.g.}, dark energy) if  the NEC (condition $ii$) is violated in the {\it compact} dimensions; that is, $\rho+p_k < 0$ for some $t$ and $y_m$.  Furthermore,  a static NEC violation is not sufficient; $\rho+p_k \ge 0$ must be time-dependent.   The proof is simple:  by choosing $A=A*$ in Eq.~\ref{essentials2} and fixing $\xi=0$, the last two terms are zero and, hence, $e^{-\phi} \langle e^{2 \Omega} (\rho+p_k)\rangle_A*$ is precisely equal to $\rho_{4d}(1+3w)$.  But $\rho_{4d}(1+3w)$ switches sign as the universe transitions from a  matter-dominated phase with decelerating expansion to a dark energy-dominated phase with accelerating expansion and its magnitude varies with time depending on $\rho_{4d}(z)$ and $w(z)$.

In short, violating condition $ii$ above is sufficient in principle to obtain an acceptable model, but then the NEC violating sources must:  (1) lie in the compact dimensions; and (2) vary with redshift in a manner that precisely tracks
the equation-of-state $w(z)$ as measured in the 4d effective theory, as discussed in \cite{Steinhardt:2008nk}.   Note that $w(z)$ depends on the matter energy density where the matter lies in the large dimensions, so condition (2), tracking $w(z)$, requires a non-trivial construction in which the NEC source in the compact dimensions couples to the energy density in the non-compact directions in a precise dynamical way.  As of this writing, we do not know of any working example.

\section{Conclusions}

The take-away message of this paper is that cosmic acceleration is even more difficult to incorporate
in compactified theories than considered previously.  This statement applies to both cosmic acceleration in the very early universe (inflation) and in the current universe (dark energy), though here we have focused on the latter. 
To satisfy both the  metric-based  and Swampland-based constraints,  we have shown theories must include a dynamical NEC violating component that is  inhomogeneously distributed in the compact dimensions and precisely in sync with $w(z)$ as measured in the 4d effective theory.  It remains an open challenge to find a concrete construction of this type and show that it can satisfy cosmological constraints.  

The alternative is to violate the Swampland-based and/or  the metric-based  constraints.   If one wishes to do this without abandoning string theory altogether, there are hurdles to cross. A realistic string-based model must: (1) explain low energy gauge theory; (2) stabilize moduli; and, (3) provide a nearly de Sitter vacuum to match cosmological observations of dark energy.  All solutions we know of that satisfy the first requirement utilize metrics of the RF or CRF type, in accord with the metric constraints assumed in this paper.   Examples that attempt to satisfy the remaining two requirements typically involve metrics other than RF or CRF or rely upon uncontrolled approximations or uncalculated nonperturbative effects that violate the Swampland constraints.   At present, we do not know of any approaches that would satisfy all three requirements, but we hope that identifying certain cosmological no-goes and restrictions, as done here, will suggest new promising directions in compactified theory construction or perhaps alternative string theoretic models that do not rely on compactification which can accomplish the feat. 

{\it Acknowledgements.} 
We thank D. Andriot, J. Louis, and C. Vafa for useful comments.
The work of P.J.S. is supported by the DOE grant number DEFG02-91ER40671 and by the Simons Foundation grant number 654561.

\bibliographystyle{apsrev}
\bibliography{bibl1.bib}
\end{document}